\documentclass[a4paper,11pt]{article}
\pdfoutput=1 

\usepackage{jheppub} 

\usepackage[T1]{fontenc} 
\usepackage{amsmath, float, graphicx,amssymb,multirow,pgfplots,pgfplotstable}
\usepackage{tikz}
\usepackage{slashed}
\usepackage{comment}
\usepackage[caption=false]{subfig}
\usetikzlibrary{shapes.geometric,arrows}
\usepackage{cleveref}
\usepackage{tikz-feynman}
\usepackage{titlesec}

\usepackage{graphicx}
\usepackage{dcolumn}
\usepackage{bm}
\usepackage{xspace}
\usepackage{cleveref}
\usepackage{color}

\titlespacing*{\section}
{0pt}{ 2.5ex}{2.5ex}\titlespacing*{\subsection}
{0pt}{ 2.5ex}{2.5ex}

\title{\boldmath Unraveling dark Higgs mechanism via dark photon production at an $e^+ e^-$ collider}

\author[a]{Song Li,}
\author[a,b]{Jin Min Yang,} 
\author[c]{Mengchao Zhang~\footnote{Corresponding author},}
\author[a]{Yang Zhang,}
\author[b,d]{Rui Zhu~\footnote{Corresponding author}}

\affiliation[a]{School of Physics, Henan Normal University, Xinxiang 453007, P. R.  China}
\affiliation[b]{Institute of Theoretical Physics, Chinese Academy of Sciences, Beijing 100190, P. R. China}
\affiliation[c]{Department of Physics and Siyuan Laboratory, Jinan University, Guangzhou 510632,  P. R.  China}
\affiliation[d]{School of Physical Sciences, University of Chinese Academy of Sciences, Beijing 100049, China}

\emailAdd{chunglee@htu.edu.cn}
\emailAdd{jmyang@itp.ac.cn}
\emailAdd{mczhang@jnu.edu.cn}
\emailAdd{yang.phy@foxmail.com}
\emailAdd{zhurui@itp.ac.cn}

\abstract{
In the phenomenology study of dark photon, its mass origin is usually not under concern. However, in theory construction its mass is often generated via a dark Higgs mechanism, which leads to the presence of a light (non-decoupled) dark Higgs particle. In this work, we study the impact of such a dark Higgs particle in the collider detection of the dark photon. We focus on the process of final state dark photon radiating dark Higgs, which is called dark final state radiation (FSR). Considering the effects on both the signal cross section and the distribution of the missing mass square, the invisible dark photon search at BaBar is reanalyzed and a new exclusion limit for invisible dark photon is presented. }

\begin{document}
\maketitle

\section{Introduction \label{sec:introduction} }
The existence of dark matter (DM) has been confirmed through its gravitational effects. 
However, its non-gravitational interactions have not been experimentally detected to date. 
The null result of DM detection indicates that the interaction between DM and the Standard Model (SM) particles may be very weak, which also largely excludes the natural parameter space of common WIMP (Weakly Interacting Massive Particle) models ~\cite{Abercrombie:2015wmb,PandaX-II:2016vec,LUX:2016ggv,ATLAS:2017nga,ATLAS:2013ndy,CMS:2016gox,ATLAS:2017uis}.
In recent years, an increasing number of studies have shifted focus to investigating the so-called ``dark sector'' models.
In such models, the DM generally couples to SM particles via a vector particle called dark photon (DP, labeled by $A'$) through a small kinetic mixing between DP and the SM photon~\cite{Arkani-Hamed:2008hhe,Pospelov:2008jd,Alves:2009nf,Hisano:2003ec,March-Russell:2008lng,Arkani-Hamed:2008kxc,Cirelli:2008pk,Pospelov:2007mp,Cholis:2008wq,Cholis:2008qq,Rizzo:2024bhn}, and thus make it easier to escape the limits on WIMPs. 
Consequently, experimental searches for these dark sector models often focus on the detection of DP.
The DP couples with the SM charged particles via a kinetic mixing, and thus can be detected by multiple experiments like beam-dump experiments~\cite{SHiP:2021nfo}, low energy electron-positron colliders~\cite{BaBar:2001yhh,Belle:2000cnh,BESIII:2009fln,Adinolfi:2002uk}, fixed target experiments~\cite{Andreas:2013lya,NA64:2016oww,Gninenko:2013rka,NA64:2017vtt,DarkSHINE:2022mak}, Electron-Ion Collider (EIC)~\cite{Yan:2022npz}, or forward detectors~\cite{Feng:2017uoz,FASER:2022hcn,LHCb:2008vvz}.
Current data already excluded a large region in the 2-dimensional parameter space ($m_{A'}, \varepsilon$), with $m_{A'}$ being the DP mass and $\varepsilon$ the kinetic mixing parameter~\cite{Banerjee:2019pds,Zhang:2019wnz,BaBar:2017tiz,BESIII:2017fwv,Ilten:2015hya,KLOE-2:2012lii,KLOE-2:2016ydq,Ilten:2016tkc,Merkel:2014avp,BaBar:2014zli,KLOE-2:2011hhj,LHCb:2019vmc,KLOE-2:2014qxg,NA482:2015wmo,Belle-II:2018jsg}. 

However, the detection of DP which only based on $m_{A'}$ and $\varepsilon$ might be oversimplified. 
It is generally considered that DP mass $m_{A'}$ is from the Stueckelberg mechanism. 
But Stueckelberg mechanism is only valid when the dark Higgs (labeled by $s$) is much heavier than DP, i.e. $\lambda \gg g'$ ($\lambda$ and $g'$ are dark Higgs self-coupling and dark $U(1)'$ gauge coupling, respectively).
Due to the perturbative unitarity bound~\cite{Lee:1977yc,Lee:1977eg}, $\lambda$ should be smaller than $4\pi$~\cite{Li:2024wqj}. 
And $g'$ is generally not too small in some model setting~\cite{Chen:2023rrl,Baldes:2017gzw,Graesser:2011wi,Pospelov:2007mp,Ibe:2019ena}. 
So, $\lambda \gg g'$ is generally invalid and we can not treat dark Higgs $s$ to be decoupled. 

There are already some works in the literature that study the DP and dark Higgs at collider simultaneously~\cite{Batell:2009yf,Cheung:2024oxh,Belle-II:2022jyy,KLOE-2:2015nli,BaBar:2012bkw,Jaegle:2015fme}. 
All of these studies rely on the assumption that DP decays mainly to lepton pair or charged mesons, i.e. they only consider visible DP.
Furthermore, most of these studies consider dark Higgsstrahlung process $e^+ e^- \to A' s$ at electron-position colliders. 
Different with previous studies, in this work we will analyze the case where both the DP and dark Higgs are invisible. 
Certainly in this invisible case the dark Higgsstrahlung process $e^+ e^- \to A' s$ can not be used to search for dark sector, because all the final state particle will not leave any trace on detectors. 
Instead, in this work we will consider dark radiation process $e^+ e^- \to \gamma A' s$ with the extra dark Higgs in the final state coming from a dark final state radiation (FSR) process $A'\to A' s$.
To our knowledge, this is the first time that the process $e^+ e^- \to \gamma A' s$ being studied. 

For invisible DP detection at colliders, the signal is generally the so-called missing mass square $M^2_X$. 
We will show that dark FSR process $A'\to A' s$ changes the distribution of $M^2_X$. 
Furthermore, when the center-of-mass collision energy $\sqrt{s}$ is much larger than $m_{A'}$ and $m_s$, fixed order perturbative calculation will cause collinear divergence. 
This problem can be solved by the ``dark shower'' methods~\cite{Buschmann:2015awa,Kim:2016fdv,Chen:2018uii,Chigusa:2022act}, which is similar to the familiar parton shower in the SM. 
By merging the contribution from dark shower and fixed order process, we obtain a reliable description of dark FSR process $e^+ e^- \to \gamma A' s$. 
A case study at BaBar will be given to show how such a dark FSR can affect the search of DP.

This paper is organized as follows. We briefly introduce the model in this study in Section~\ref{sec-ii}, 
and delineate how to characterize the FSR in the dark sector in Section~\ref{sec-iii}. 
In Section~\ref{sec-iv}, we formally analyze the potential impact of the dark FSR.
As a case study, in Section~\ref{sec-v}, concentrating on the BaBar collaboration's search for invisible dark photon, we quantitatively examine the effect of the dark FSR on the exclusion limit. 
This work is concluded in Section~\ref{sec:conclusion}.

\section{The dark sector model\label{sec-ii}}
The dark sector we study in this work is charged under a dark $U(1)'$ gauge symmetry which is spontaneously broken via the dark Higgs mechanism. 
The complete Lagrangian for the dark sector before symmetry breaking is ($\text{diag}(+,-,-,-)$ metric ):
\begin{equation}
\mathcal{L}_{\text{dark}} = ( D_\mu S )^\dagger D^\mu S - \frac{1}{4} F'_{\mu\nu} F'^{\mu\nu} + \mu^2 S^\dagger S - \frac{1}{4} \lambda (S^\dagger S)^2 + \bar{\chi} ( i \slashed{D} - m_\chi ) \chi - \varepsilon e J^{\text{EM}}_\mu A'^{\mu}\label{eq:lag}
\end{equation}
with $F'_{\mu\nu} = \partial_\mu A'_\nu - \partial_\nu A'_\mu$ the field strength of DP $A'$. 
$D_\mu = \partial_{\mu} - ig' A'_\mu$ is the covariant derivative and $g'$ is the dark gauge coupling. 
$S$ is the complex dark Higgs field before symmetry breaking.  
Light dark fermion $\chi$ is introduced to make DP decay invisibly. 
$\varepsilon$ and $e$ are the kinetic mixing parameter and electric charge, respectively.
$J^{\text{EM}}_\mu$ is  the SM electromagnetic current. 

After symmetry breaking, dark Higgs field obtain vacuum expectation value (VEV) $v={2\mu}/{\sqrt{\lambda}}$ and can be expanded as $S=\frac{1}{\sqrt{2}}(v+s+ia)$. 
The masses of dark Higgs $s$ and dark photon $A'$ are given by: 
\begin{eqnarray}
m_s^2 = \frac{1}{2} \lambda v^2 \ , \ m^2_{A'} = g'^2 v^2
\end{eqnarray}

To simplify our analysis, in the rest of this paper we will assume $m_s = m_{A'}$. 
This assumption does not affect the conclusion much. 
The mass of dark fermion $\chi$ is simply taken to be much smaller than $m_{A'}$, which makes the invisible decay channel $A'\to \bar{\chi}\chi$ to be dominant. 
We further assume the coupling of Higgs portal $(S^\dagger S)(H^\dagger H)$ ($H$ is the SM Higgs) to be negligible, and thus dark Higgs $s$ can only decay via off-shell dark photons ($s\to A'^\ast A'^\ast\to \bar{\chi}\chi \bar{\chi}\chi $) which also makes dark Higgs invisible.

Compared with the conventional DP model with 2 input parameters $\{ \varepsilon , \  m_{A'} \}$, our dark Higgs model has 3 input parameters $\{ \varepsilon , \ m_{A'}(=m_s) , \ \alpha' \}$. Here $\alpha' = g'^2/{4\pi}$ is the dark fine structure constant.  

\section{Final-state radiation (FSR) in the dark sector\label{sec-iii}}

The DP can be generated by high energy charged particles via the kinetic mixing. 
Because the mixing parameter $\varepsilon$ is very small, we consider the process with only one DP in the final state. 
In our dark sector model with a Higgs mechanism, there are 3-point couplings between dark Higgs and DP, such as $A'_\mu A'^\mu s$ and $ A'^\mu  ( a\! \stackrel{\leftrightarrow}{\partial_\mu}\!  s ) $. 
So it is possible for a final-state DP to emit a dark Higgs if $\alpha'$ is not too small. 
This dark Higgs emitting process, $A'\to A' s$, will be called the dark FSR in the rest of this paper. 

\begin{figure}[htb]
	\centering
	\includegraphics[width=8.5cm]{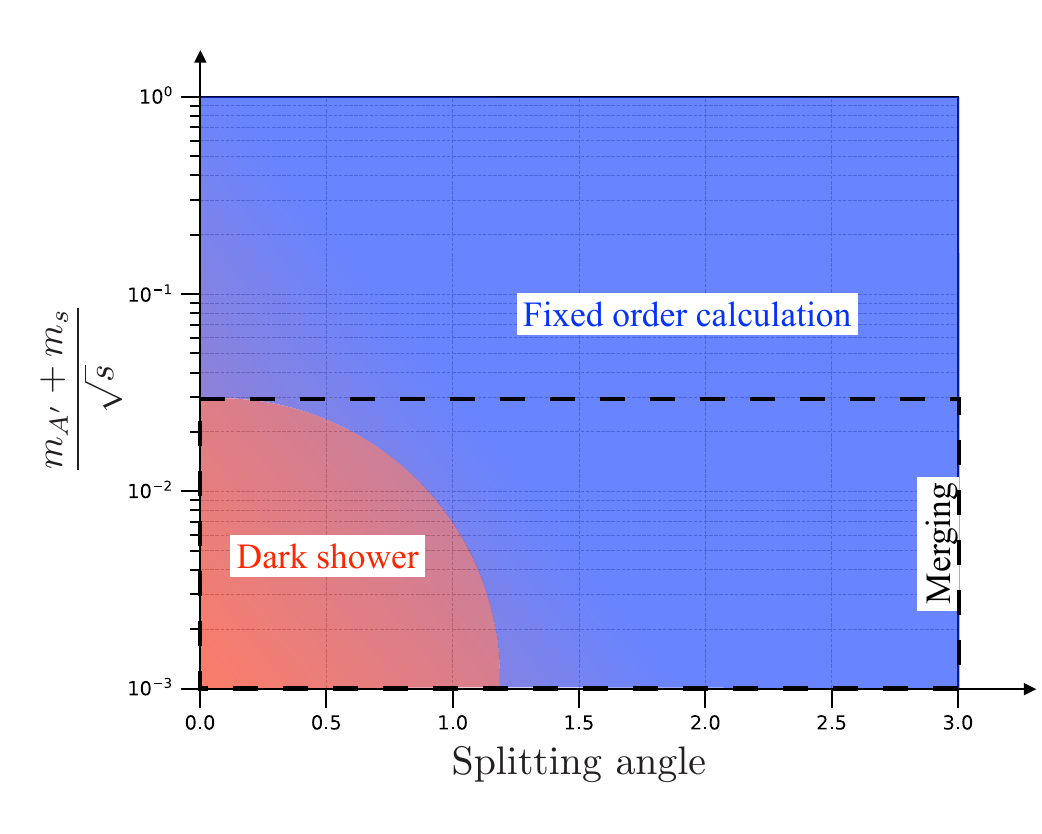}
  \vspace{-.5cm}
	\caption{Illustration for the method to deal with the dark FSR. 
    }
\label{fig:vs}
\end{figure}

To deal with the dark FSR, we adopt the method as illustrated in the diagrammatic sketch Fig.~\ref{fig:vs}. 
Comparing to the center-of-mass energy $\sqrt{s}$ for a certain collider, if $(m_{A'} + m_s)$ is not too small, the fixed-order calculation is valid.
But when $(m_{A'} + m_s) \ll \sqrt{s}$, the description of collinear radiation by the fixed-order calculation will show unphysical enhancement or divergence. 
Similar to parton shower~\cite{Bengtsson:1986et,Chen:2016wkt,Collins:1984kg,Collins:1989gx,Sudakov:1954sw} in the SM, we use the so-called dark shower~\cite{Buschmann:2015awa,Kim:2016fdv,Chen:2018uii,Chigusa:2022act} to re-sum multiple radiation processes and describe this collinear radiation.
Then the result of dark shower should be merged with the fixed-order result, which describes the non-collinear phase space, to give a complete events distribution.
Detailed discussions about dark shower and the merging method are given in the following. 

\subsection{Dark shower}
The differential splitting function $d\mathcal{P}(A'\to s+A')$ quantifies the probability of ``finding'' a dark Higgs $s$ within a DP $A'$. If a final-state DP emits a soft collinear dark Higgs, since the propagator $A'$ remains approximately on-shell, the differential cross section with dark FSR can be factorized as: 
\begin{eqnarray}
d\sigma(X\to Y + A' + s) \simeq d\sigma(X\to Y + A' ) \times  d\mathcal{P}(A'\to  s + A') 
\end{eqnarray}

$d\mathcal{P}(A'\to  s + A')$ is  typically expressed as a function of the energy fraction  of $s$ relative to mother particle $A'$ (denoted by $z$) and the transverse momentum of $s$ perpendicular to the mother $A'$ (denoted  by $p_T$). The calculation details of $d\mathcal{P}(A'\to  s + A')$ are given in Appendix~\ref{appA}. In the following analysis, we focus exclusively on the leading-power results and the corresponding splitting kernels are
\begin{gather}
P_{A'_{T} \to s + A'_L}(z)
= z\bar{z},\\
P_{A'_{L} \to s + A'_T}(z)
=\frac{2z}{\bar{z}},
\end{gather}
in which $\bar{z} \equiv 1-z$. 
Unlike conventional parton showers, at the leading-power, the radiation of the dark Higgs alters the polarization of the dark photon, converting transverse dark photons to longitudinal ones, and vice versa. 

With these splitting kernels, the dark shower in our model can be realized numerically by the veto algorithm~\cite{Hoeche:2009xc,Lonnblad:2012hz,Mrenna:2016sih} implemented in \textsc{Pythia8.3}~\cite{Bierlich:2022pfr}.  
Specifically, we modified the code originally developed for the Hidden Valley model \cite{Carloni:2010tw, Carloni:2011kk}, implemented alternating calls to the splitting kernels of $A'_T \to s A'_L$ and $ A'_L \to s A'_T$ to model the timelike final-state shower of DP, adopting a transverse-momentum-ordered evolution scheme. Since only transversely polarized DPs couple to the electromagnetic current, we assume all DPs are purely transverse prior to the shower evolution at the high-energy limit.

\begin{figure}[!t]
	\centering
	\includegraphics[width=8.5cm]{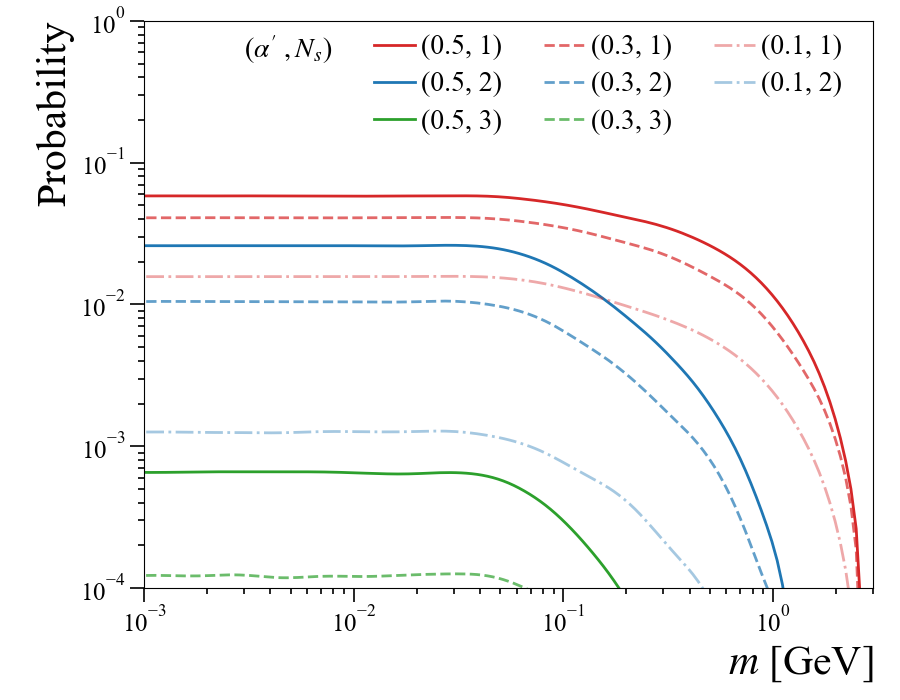}
  \vspace{-.2cm}
  \captionsetup{justification=raggedright, singlelinecheck=false}
	\caption{The probability for a final-state DP to radiate $N_s$ dark Higgs bosons given by the parton shower simulation. Solid, dashed and dash-dotted curves correspond to $\alpha' = 0.5$, $0.3$ and $0.1$, respectively. Red, blue and green curves represent $N_s = 1$, $2$ and $3$, respectively. }
\label{fig:Ns}
\end{figure}

To illustrate the effect of dark shower, we  simulate the probability of radiating $N_s$ dark Higgs as a function of mass $m$ ($m = m_{A'} = m_{s}$). The model is implemented  with \textsc{FeynRules} \cite{Alloul:2013bka} and tree-level $e^+e^- \to \gamma A'$ events are generated by \textsc{MadGraph5} \cite{Alwall:2014hca, Frederix:2018nkq}. Then turn to \textsc{Pythia8.3} \cite{Bierlich:2022pfr} to simulate the dark shower. And the result is shown in Fig.~\ref{fig:Ns}.
If $m$ is much smaller than central energy $\sqrt{s} = 10$ GeV, then the probability of radiating a dark Higgs is about $6\%$ for $\alpha'=0.5$.
 But the probability of radiating three or more dark Higgs bosons is always below $0.1\%$. 
The radiating probability exhibits negligible $m$-dependence for $m\ll \sqrt{s}$ and decreases as $m$ above a threshold.

\subsection{Merging in the dark sector}  
\begin{figure}
	\begin{center}
	\includegraphics[width=8.5cm]{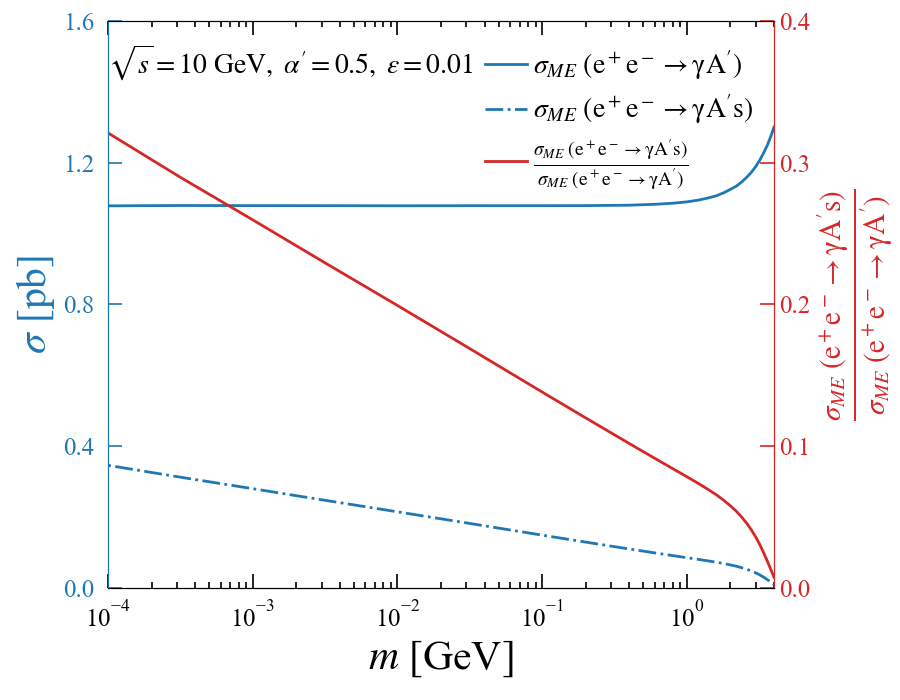}
    \end{center}
  \vspace{-.8cm}
  
  \captionsetup{justification=raggedright, singlelinecheck=false}
	\caption{The tree-level cross sections (with  $p^T_{\gamma}>1$ GeV) and their ratio as functions of $m = m_{A'} = m_{s}$. The solid and dash-dotted blue lines are the cross sections for the $e^+ e^- \to \gamma A' s$ and $e^+ e^- \to \gamma A'$ processes, respectively, while the red line shows their ratio. }
\label{fig:mgcxs}
\end{figure}

Consider the dark photon production associating with a photon on an $e^+~e^-$ collider, i.e. $e^+e^-\to\gamma A'$. Fig.~\ref{fig:mgcxs} shows the cross sections of $e^+e^-\to\gamma A'$ and $e^+e^-\to\gamma A' s$ at tree level, assuming both the dark Higgs and DP masses equal to $m$, and taking $\sqrt{s} = 10~\text{GeV}$, $\varepsilon = 0.01$ and $\alpha' = 0.5$. (Calculated with \textsc{MadGraph5} \cite{Alwall:2014hca, Frederix:2018nkq} and the basic cut $p^{\gamma}_{T} > 1~\text{GeV}$ is taken.) 
The ratio $\sigma_{\text{ME}}(e^+e^-\to\gamma A' s)/\sigma_{\text{ME}}(e^+e^-\to\gamma A')$, which is independent of the value of $\varepsilon$, is also given.

For a large $m$, it is reasonable to take the fixed-order matrix element (ME) calculation of the three-final-states process as the leading-order contribution from the dark FSR:
\begin{eqnarray}
\sigma_{\text{FSR}}=\sigma_{\text{ME}}(e^+e^-\to\gamma A' s)~~\text{(for~large~$m$).}
\label{eq:cxsh}
\end{eqnarray}

When $m \ll \sqrt{s}$, the physical cross section should be insensitive to the variation of $m$, whereas from Fig.~\ref{fig:mgcxs}, tree-level $\sigma_{\text{ME}}(e^+e^-\to\gamma A' s)$ increases significantly as $m$ decreases. This occurs because in the $m \to 0$ limit, the collinear radiation introduces infrared (IR) divergence.

In the calculation of total cross section, this IR divergence can be canceled by virtual correction~\cite{Dev:2024twk}, as promised by Kinoshita-Lee-Nauenberg theorem~\cite{Kinoshita:1962ur,Lee:1964is}. 
But the physical final state distribution can not be obtained by considering virtual correction. 
In the following, for small $m$, the merging method is introduced to the dark sector to obtain both the distributions and the cross section without divergence. 

 \begin{figure}[htb]
	\centering
	\includegraphics[width=8.5cm]{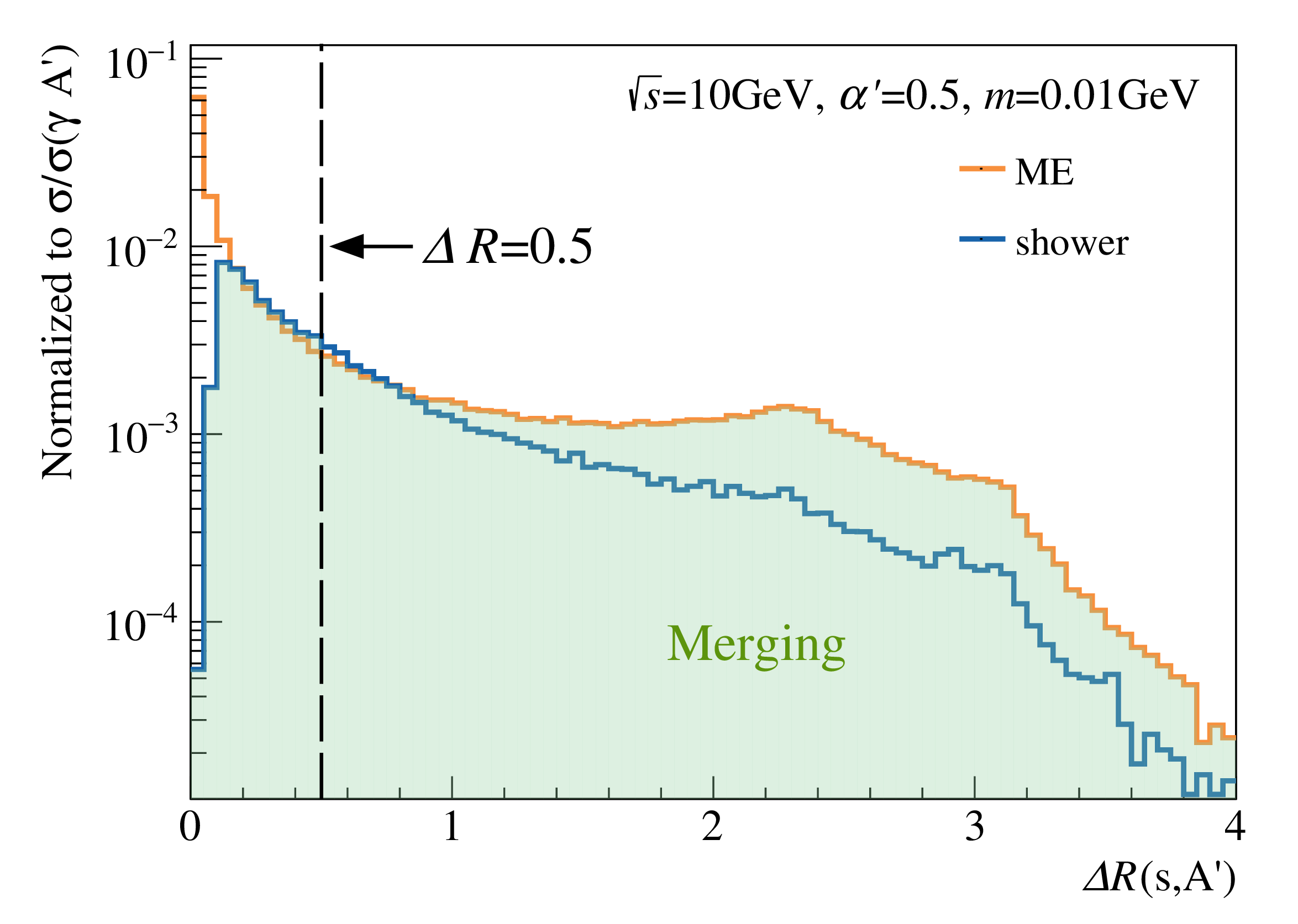}
  \vspace{-.2cm}
  \captionsetup{justification=raggedright, singlelinecheck=false}
	\caption{The $\Delta R$ distributions and a schematic illustration of the merging procedure. The orange line represents the hard scattering process $e^+ e^- \to \gamma A' s$. The blue histogram is filled with  showered events of $e^+ e^- \to \gamma A'$ with at least one dark Higgs boson in the final state. And both are normalized to the ratio of their respective cross sections to $\sigma_\text{ME}(e^+ e^- \to \gamma A')$. The green shaded region represents the distribution after merging.}
\label{fig:merging}
\end{figure}

After the dark shower, the MC events not necessarily contain dark Higgs in their final-state, i.e. the dark FSR might not happen. The probability for dark shower to radiate at least one dark Higgs (sh-wiDH) is denoted as $\beta$ (as shown in Fig.~\ref{fig:Ns}). Then the corresponding cross section is: 
\begin{eqnarray}
\sigma_{\text{sh-wiDH}}=\frac{\beta}{1-\beta}\times \sigma_{\text{ME}}(e^+ e^- \to \gamma A'). 
\end{eqnarray}

Then combine the results of shower and the fixed-order calculation. Partition the phase space using $\Delta R \equiv \sqrt{(\Delta\eta)^2 + (\Delta\phi)^2}$ between $A'$ and $s$. For example, at $m=0.01$ GeV and $\alpha'=0.5$, Fig.~\ref{fig:merging} displays the $\Delta R$ distribution in which the blue line is for PYTHIA-simulated shower events (aquiring at least one dark Higgs in the final-state) and the orange line is for MadGraph-generated fixed-order events ($e^+e^- \to \gamma A' s $). Using $\Delta R_{\text{cut}}=0.5$ as the separatrix, the cross section (for small $m$) contributed by the dark FSR process is
\begin{eqnarray}
\sigma_{\text{FSR}} = P_{\text{sh-wiDH}}(\Delta R<\Delta R_{\text{cut}})\sigma_{\text{sh-wiDH}}  + P_{\text{ME}}(\Delta R>\Delta R_{\text{cut}})\sigma_{\text{ME}}(e^+ e^- \to \gamma A' s)
\label{eq:cxsl}
\end{eqnarray}
where $P_{\text{sh-wiDH}}$ and $P_{\text{ME}}$ denote the probabilities for shower and fixed-order events to satisfy the respective $\Delta R$ criteria.

By weighting the Monte Carlo (MC) events with their respective cross-sections, the correct distributions can be obtained.

\section{Impact of the dark FSR \label{sec-iv}}
In our model both the DP and the dark Higgs decay invisibly, thus the process $e^+ e^- \to \gamma A' s$ presents the same signature as $e^+ e^- \to \gamma A'$, i.e. mono-photon plus missing energy. So the signal cross section enhances from
\begin{eqnarray}
\sigma_{\text{no-FSR}}=\sigma_{\text{ME}}(e^+ e^- \to \gamma A')
\label{eq:cxssgn}
\end{eqnarray}
to
\begin{eqnarray}
\sigma_{\text{tot}}=\sigma_{\text{no-FSR}}+\sigma_{\text{FSR}}.
\label{eq:cxssgn}
\end{eqnarray}

In addition, considering the dark FSR, the distributions of kinematic variables - such as the squared missing mass ($M_X^2 = s - 2E_{\gamma}\sqrt{s}$, with $E_{\gamma}$ being the photon energy in the center-of-mass frame) - are modified.

Without dark FSR, the theoretical distribution of $M_X^2$ for the signal is nearly a delta function centered at the DP mass $m_{A'}$ (if the decay width of DP is very small):
\begin{eqnarray}
f^{\text{th}}_{\text{no-FSR}}(M_X^2) = \delta(M_X^2 - m_{A'}^2).
\end{eqnarray}
When dark FSR is included, the total theoretical $M_X^2$ distribution is a weighted summation
\begin{eqnarray}
f^{\text{th}}_{\text{tot}}(M_X^2) = \lambda f^{\text{th}}_{\text{no-FSR}}(M_X^2) + (1-\lambda) f^{\text{th}}_{\text{FSR}}(M_X^2),
\end{eqnarray}
in which
\begin{eqnarray}
\lambda = \frac{\sigma_{\text{no-FSR}}}{\sigma_{\text{no-FSR}} + \sigma_{\text{FSR}}}.
\end{eqnarray}
represents the fraction of events contributed by the two-final-states process. $f^{\text{th}}_{\text{FSR}}(M_X^2)$ is the distribution of events from the FSR contribution and can be approximated by filling the histogram of $M_X^2$ with MC events and performing a linear interpolation. Note that in the low-mass region, the merged MC events are used.

The experimentally observed signal distribution $f^{\text{ex}}_{\text{tot}}(M_X^2)$ results from the convolution of this theoretical distribution $f^{\text{th}}_{\text{tot}}(M_X^2)$ with the detector resolution function $f^{\text{re}}(M_X^2)$,
\begin{eqnarray}
f^{\text{ex}}_{\text{tot}}(M_X^2) &=& f^{\text{th}}_{\text{tot}}(M_X^2) \otimes f^{\text{re}}(M_X^2) .
\end{eqnarray}
Due to the presence of $f^{\text{th}}_{\text{FSR}}(M_X^2)$ contributed by the dark FSR, both $f^{\text{th}}_{\text{tot}}(M_X^2)$ and $f^{\text{ex}}_{\text{tot}}(M_X^2)$ 
are flattened.

\section{Recast BaBar analysis with dark FSR effects \label{sec-v}} 
For the $e^+ e^- \to \gamma A'$ channel with $A' \to \text{invisible}$, the BaBar collaboration has analyzed data collected at the PEP-II B-Factory near the $\Upsilon(2S)$, $\Upsilon(3S)$, and $\Upsilon(4S)$ resonance peaks ($\sqrt{s}\sim 10$ GeV), with an integrated luminosity of 53 fb$^{-1}$\cite{BaBar:2017tiz}. They derived the exclusion limit on the mixing parameter $\varepsilon$ via a maximum likelihood fit to the distribution of $M_X^2$.
In this section, we quantify the specific impact of the dark FSR on BaBar's exclusion limit under the assumption $\alpha'=0.5$ and for $m$ in the range $[0.001, 2]$ GeV.

Let $N_0$ denotes the original signal count upper limit without dark FSR, corresponding to a mixing parameter upper limit $\varepsilon_0$. With dark FSR considered, they become $N'$ and $\varepsilon'$, respectively. Since the signal count is proportional to the cross section, and the cross section scales quadratically with the mixing parameter, we can seperate the impact of the dark FSR into two parts 
\begin{eqnarray}
\frac{\varepsilon'^2}{\varepsilon_0^2} = \frac{N'}{N_0} \frac{\sigma_{\text{no-FSR}}(\varepsilon=1)}{\sigma_{\text{tot}}(\varepsilon=1)} = S_{\text{fit}} \cdot S_{\text{cxs}}.
\label{eq:limit}
\end{eqnarray} 
Here, $S_{\text{fit}} \equiv N'/N_0$ represents the correction factor due to the altered signal shape affecting the fit, while $S_{\text{cxs}} \equiv \sigma_{\text{no-FSR}} / \sigma_{\text{tot}}$ is the correction factor from the cross section increase.  

The distribution of $M_X^2$ influences the confidence intervals by affecting the likelihood function under the signal-plus-background (S+B) hypothesis. The analysis from BaBar has 6 signal regions for the low-mass situation, and for simplicity, we only focus on the region near the $\Upsilon(3S)$ resonance satisfying the $\mathcal{R}'_L$ selection criteria, which contains 129 events, more than the sum of the other five regions. We get the distributions by MC simulation, construct the profile-likelihood ratio test statistic, then perform hypothesis test inversion using \textsc{RooStats} \cite{Moneta:2010pm, Cowan:2010js} to obtain the $90\%$ confidence level (C.L.) upper limits (based on the CLs method) on the signal yield $N_0$ (without dark FSR) and $N'$ (with dark FSR). Taking $m=0.01$ GeV as an example, the detailed process is given in Appendix~\ref{appB}. Repeating for different $m$, the factor $S_{\text{fit}} \equiv N'/N_0$ is always in the range of 1.03 to 1.04.

\begin{figure}[htb]
	\centering
	\includegraphics[width=8.5cm]{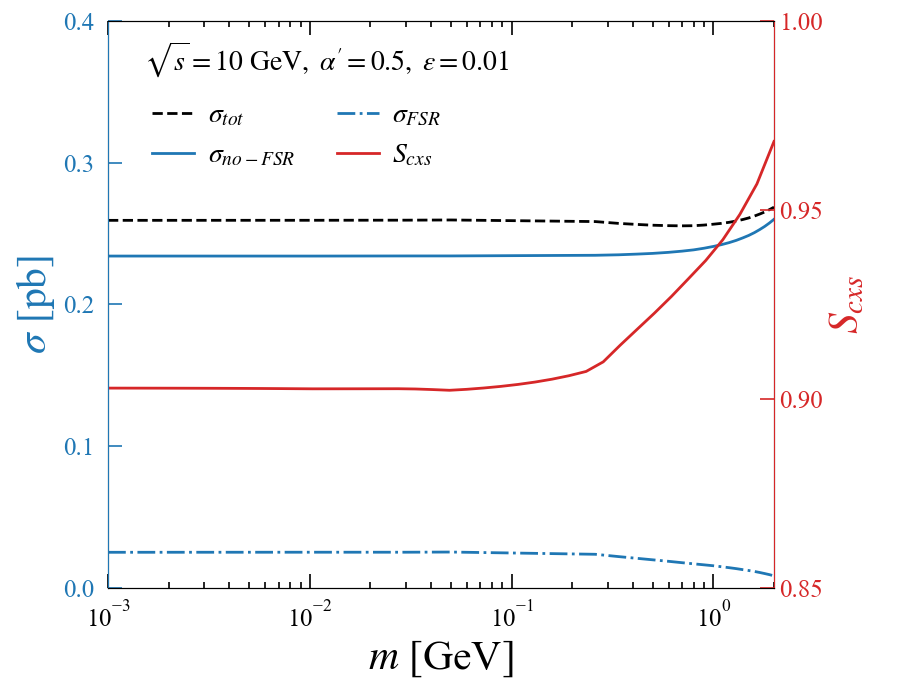}
  \vspace{-.2cm}
\captionsetup{justification=raggedright, singlelinecheck=false}
	\caption{The signal cross sections as functions of $m = m_{A'} = m_{s}$. The black dashed curve represents the cross section incorporating dark FSR. The solid and dash-dotted blue lines are for the components without or with dark Higgs bosons in final state, respectively. The red line represents the correction factor $S_{\text{cxs}} \equiv \sigma_{\text{no-FSR}} / \sigma_{\text{tot}}$ introduced to $\varepsilon^2$ due to the cross section variation.}
\label{fig:cxsmerge}
\end{figure}

The cross sections assuming $\varepsilon=0.01$ is given in Fig.~\ref{fig:cxsmerge}, requiring   $E_{\gamma} > 3$ GeV, $|\cos\theta_{\gamma}| < 0.6$, and $-4 < M_X^2 < 36$ GeV$^2$, where $E_{\gamma}$ and $\cos\theta_{\gamma}$ represent energy and polar angle of photon in the center-of-mass frame.
These selection criteria are consistent with Ref.~\cite{BaBar:2017tiz}.
Note that the cut efficiency is also influenced by the PDF of $M_X^2$ for signal. For the cross section $\sigma_{\text{FSR}}$ contributed by the dark FSR, we utilize the merging result for $m < 0.05$ GeV  and ME result for $m > 0.25$ GeV, with a smooth interpolation applied in the intermediate mass region.
 $S_{\text{cxs}}$ is about 0.90 for $m < 0.1$ GeV, and increases as $m$ increases above a threshold.

\begin{figure}[htb]
	\centering
	\includegraphics[width=8.5cm]{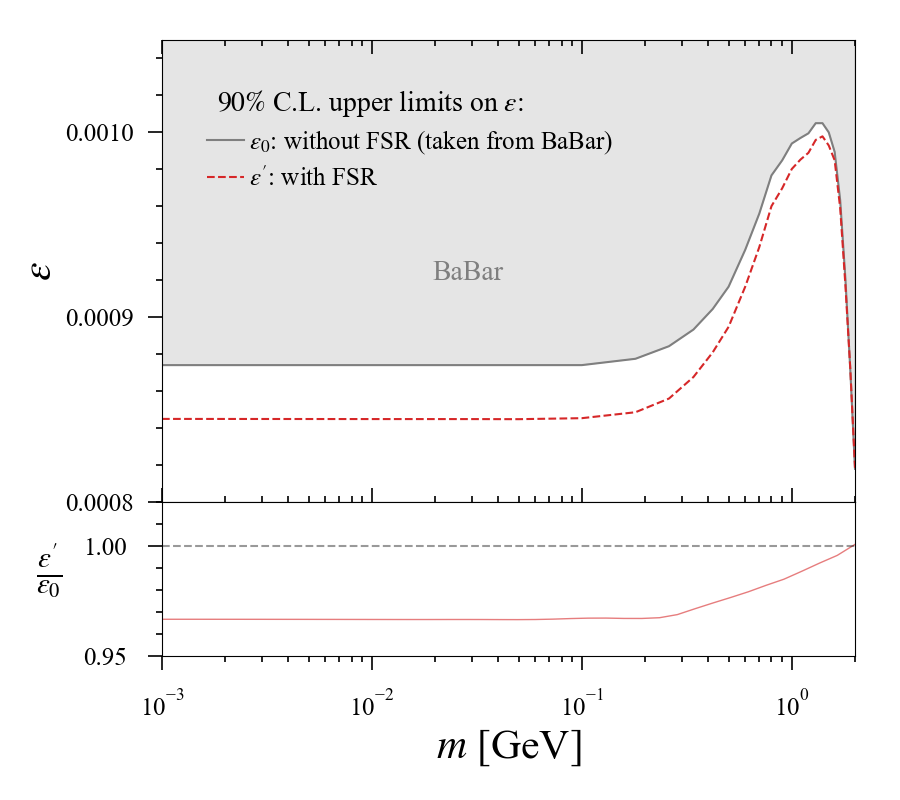}
  \vspace{-.5cm}
  \captionsetup{justification=raggedright, singlelinecheck=false}
	\caption{Upper limits on the mixing parameter $\varepsilon$ at 90\% C.L., comparing scenarios with and without dark FSR. The gray curve corresponds to the profile-likelihood limit from BaBar\cite{BaBar:2017tiz}. The red dashed curve shows the exclusion limit incorporating dark FSR effects from the dark Higgs mechanism, under the assumption $\alpha'=0.5$ and $m = m_{A'} = m_{s}$. The red line in the lower panel gives the ratio of the new limit to the BaBar limit.}
\label{fig:limit}
\end{figure}

Following Eq.~\ref{eq:limit}, the exclusion limit $\varepsilon'$ incorporating dark FSR effect is derived and shown as the red dashed curve in Fig.~\ref{fig:limit}. This limit lies slightly below the limit $\varepsilon_0$ obtained without dark FSR by BaBar. For $m < 0.2$ GeV, $\varepsilon'/\varepsilon_0 \approx 0.97$. Then as $m$ increases, $\varepsilon'/\varepsilon_0$ rises gradually. 
From the perspective of systematic error, the relative uncertainty of the upper limit at $90\%$ C.L. of $\varepsilon$ introduced by the dark Higgs mechanism is estimated as $|\varepsilon_0-\varepsilon'|/\varepsilon_0$, and its maximum value is about $3.4\%$.

\section{Conclusions \label{sec:conclusion} }
In this work, we investigated how the emergence of the dark Higgs particle would affect the collider detection of dark photons.
A dark sector model with a spontaneously broken $U(1)'$ is given. 
In this model, it is possible for a high energy dark photon to emit a dark Higgs via dark FSR $A' \to A' s$.
For the $e^+ e^- \to \gamma A'$ channel searching for invisible DPs, we analyzed the impact of the dark FSR from two aspects: the increase of cross section at a fixed $\varepsilon$ which strengthen the exclusion limit;
 and the broadening of the signal's distribution which reduce detection sensitivity. 
Through a recast of the BaBar experiment, we derived a new exclusion limit on the kinetic mixing parameter $\varepsilon$. Since the cross-section enhancement effect dominates, the new limit is moderately stronger than the original result.

\addcontentsline{toc}{section}{Acknowledgments}
\acknowledgments
We appreciate helpful discussions with Alessandra Filippi, Haitao Li, Yang Ma, Lianyou Shan, Brian Shuve, Lei Wu,  Yongchao Zhang and Yu Zhang. This work is supported by the National Natural Science Foundation of China under Grant Nos. 12335005, 12105248, 12105118 and 11947118, by the Peng-Huan-Wu Theoretical Physics Innovation Center funded by the National Natural Science Foundation of China (Grant No.12447101) and by the Research Fund for PI from Henan Normal University under Grant No. 5101029470335. 

\appendix
\section{Derivation of splitting functions}  \label{appA}
\hspace{\parindent} In this part, we give details on the calculation of the splitting functions for the process $A' \to s A'$, including Feynman rules and the setting of kinematics.

\subsection{Feynman rules\label{sec:th2}}
\hspace{\parindent} The Lagrangian is given in Eq.~\ref{eq:lag}. Adopting the Goldstone Equivalence Gauge following Ref.~\cite{Chen:2018uii}, the contribution of the Goldstone component in the longitudinally polarized dark photon can be disentangled, with the gauge-fixing term specified as
\begin{equation}
    \mathcal{L}_{fix}=-\frac{1}{2\xi}(n\cdot A)(n^*\cdot A),(\xi\to0),
\end{equation}
in which
\begin{equation}
    n^\mu(k)\equiv(1,-\hat{\bold{k}}).
\end{equation}

The masses of $s$ and $A'$ are
\begin{equation}
    m_s=\sqrt{\frac{\lambda}{2}} v,\\
    ~m_{A'}=g'v
\end{equation}

The coupling terms of $A'A's$, $aas$ and $A'as$ are $g'^2 v A'_\mu A'^\mu s$, $-\frac{1}{4}\lambda v a^2 s$ and $g' \left( a \partial_\mu s - s \partial_\mu a \right) A'^\mu$ respectively, from which the following Feynman rules are derived:

\begin{eqnarray}
\begin{tikzpicture}[scale=0.7, transform shape ,baseline=(a)]
    \begin{feynman}
    \vertex(a){\(A'\)};
    \vertex[right=of a](b);
    \vertex[above right=of b](c){\(s\)};
    \vertex[below right=of b](d){\(A'\)};
    \diagram*{
    (a)--[boson](b)--[boson](d),
    (b)--[scalar](c),
    };
    \end{feynman}
    \end{tikzpicture}
    =2ig'^2v \ , \ 
    \begin{tikzpicture}[scale=0.7, transform shape ,baseline=(a)]
    \begin{feynman}
    \vertex(a){\(a\)};
    \vertex[right=of a](b);
    \vertex[above right=of b](c){\(s\)};
    \vertex[below right=of b](d){\(a\)};
    \diagram*{
    (a)--[scalar](b)--[scalar](d),
    (b)--[scalar](c),
    };
    \end{feynman}
    \end{tikzpicture}
    =-\frac{i}{2}\lambda v \ , \ 
    \begin{tikzpicture}[scale=0.7, transform shape ,baseline=(a)]
    \begin{feynman}
    \vertex(a){\(A'^\mu\)};
    \vertex[right=of a](b);
    \vertex[above right=of b](c){\(s\)};
    \vertex[below right=of b](d){\(a\)};
    \diagram*{
    (a)--[boson](b),
    (b)--[scalar,momentum=\(p_s\)](c),
    (b)--[scalar,momentum'=\(p_a\)](d),
    };
    \end{feynman}
    \end{tikzpicture}
    =g' (p_{a}^{\mu}-p_{s}^{\mu}) 
\end{eqnarray}

The $a-A'$ mixing term $-g' v A'^\mu \partial_\mu a = -m_{A'} A'^\mu \partial_\mu a$ leads to  
\begin{equation}
    \begin{tikzpicture}[scale=0.5,baseline=(b)]
    \begin{feynman}
    \vertex(a){\(A'^\mu\)};
    \vertex[right=of a](b);
    \vertex[right=of b](c){\(a\)};
    \diagram*{
    (a)--[boson](b)--[scalar,momentum=\(p_a\)](c),
    };
    \end{feynman}
    \end{tikzpicture}
    =m_{A'}p_{a}^{\mu}.
\end{equation}

For the longitudinal $A'$, which contains both the Goldstone component $a$ and the gauge component $A'_n$, the relative phase between them must be considered. In the light of the Ward identity, a phase factor $-i$ is assigned to the initial-state $a$, while a phase factor $i$ is introduced for the final-state $a$.
\subsection{Kinematics}
\begin{figure}[!t]
	\centering
	\includegraphics[width=9.5cm]{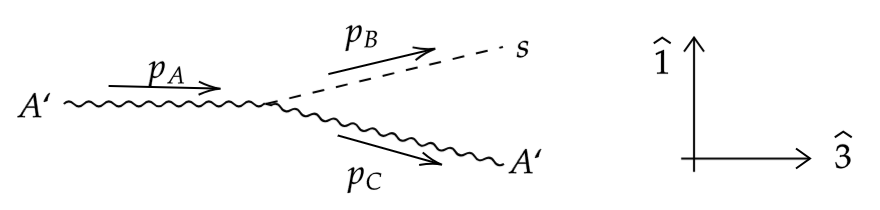}
  \vspace{-.5cm}
	\caption{The setting of kinematic variables.}
\label{fig:col}
\end{figure}

\hspace{\parindent} In the high-energy limit, we establish the coordinate system as illustrated in Fig.~\ref{fig:col} and define kinematic variables accurate up to $\mathcal{O}(p_T)$,
\begin{equation}
    \begin{split}
    p_A&=(p,0,0,p),\\
    p_B&=\left(zp,p_T,0,zp\right),\\
    p_C&=\left((1-z)p,-p_T,0,(1-z)p\right).
    \end{split}
\end{equation}
The transversely polarized vectors are
\begin{equation} 
    \begin{split}   
      \epsilon^\mu_{A+}=\frac{1}{\sqrt{2}}({0,1,i,0})&,\epsilon^\mu_{A-}=\frac{1}{\sqrt{2}}({0,1,-i,0}),\\
    \epsilon^\mu_{C+}=\frac{1}{\sqrt{2}}\left({0,1,i,\frac{p_T}{(1-z)p}}\right)&,\epsilon^\mu_{C-}=\frac{1}{\sqrt{2}}\left({0,1,-i,\frac{p_T}{(1-z)p}}\right).
    \end{split}
\end{equation}
And in the Goldstone Equivalence Gauge, the longitudinally polarized vector is
\begin{equation}
    \epsilon_L^\mu = \frac{k^\mu}{m_{A'}}-\frac{m_{A'}}{n\cdot k}n^\mu.
\end{equation}
Here, the term $\frac{k^\mu}{m_{A'}} $ corresponds to the Goldstone component, while the other term is defined as $\epsilon_n^\mu \equiv -\frac{m_{A'}}{n\cdot k}n^\mu$ and can be specifically written as
\begin{gather}
\epsilon_{An}^\mu=\frac{m_{A'}}{2p}(-1,0,0,1),
\epsilon_{Cn}^\mu=\frac{m_{A'}}{2(1-z)p}\left(-1,-\frac{p_T}{(1-z)p},0,1\right).
\end{gather}

\subsection{The differential splitting function\label{sec:th1}}
\hspace{\parindent} The differential splitting function $d\mathcal{P}(A\to B+C)$ can be derived by computing the invariant matrix element of the $1 \to 2$ process, 
\begin{equation}\label{eq:dsf}
    \frac{d\mathcal{P}}{dz dp_T^2}(A \to B + C)
=\frac{1}{16\pi^2}z\bar{z}\frac{|\mathcal{M}(A \to B + C)|^2}{p_T^4},
\end{equation}
where $\bar{z} = 1 - z$. Since $d\mathcal{P}/(dzdp_T^2)$ is always proportional to the coupling constant $\alpha'$ and $1/p_T^2$, the remaining factor is defined as the splitting kernel $P_{A\to B + C}(z)$,
\begin{equation}
    \frac{d\mathcal{P}}{dz dp_T^2}(A \to B + C)
=\frac{\alpha}{2\pi}P_{A\to B+C}(z)\frac{1}{p_T^2}.
\end{equation}
Notably, after spontaneous symmetry breaking, the splitting function requires modifications to account for the acquired particle masses \cite{Chen:2016wkt,Chen:2018uii},
\begin{equation}
    \frac{d\mathcal{P}}{dz dp_T^2}\to \frac{p_T^4}{\tilde{p}_T^4}\frac{d\mathcal{P}}{dz dp_T^2},
~~\tilde{p}_T^2\equiv p_T^2+\bar{z}m_B^2+zm_C^2-z\bar{z}m_A^2.
\end{equation}

Substituting the Feynman rules and kinematic variables into Eq.~\ref{eq:dsf}, we get the splitting functions. The leading-power contribution ($\propto \frac{p_T^2}{\tilde{p}_T^4}$) yields
\begin{gather}
    \frac{d\mathcal{P}}{dz dk_T^2}(A'_{T} \to s + A'_L)
=\frac{\alpha'}{2\pi} z\bar{z}\frac{p_T^2}{\tilde{p}_T^4},\\
\frac{d\mathcal{P}}{dz dk_T^2}(A'_{L} \to s + A'_T)
=\frac{\alpha'}{2\pi} \frac{2z}{\bar{z}}\frac{p_T^2}{\tilde{p}_T^4}.
\end{gather}
For the leading power, the radiation of the dark Higgs alters the polarization of the dark photon.
Furthermore, in the high-energy limit, the amplitudes $\mathcal{M}(A'_T \to s + A'_n)$ and $\mathcal{M}(A'_n \to s + A'_T)$, which scale as $\frac{m_{A'}^2}{p^2}$, are suppressed. Consequently, the leading-power contribution arises solely from the Goldstone component in the longitudinally polarized dark photon, consistent with the Goldstone equivalence theorem. 

Splitting processes preserving the polarization of the dark photon contribute at next-to-leading power ($\propto \frac{m_{A'}^2}{\tilde{p}_T^4}$):
\begin{gather}
    \frac{d\mathcal{P}}{dz dk_T^2}(A'_T \to s + A'_T)
=\frac{\alpha'}{2\pi} 2z\bar{z}\frac{m_{A'}^2}{\tilde{p}_T^4},\\
\frac{d\mathcal{P}}{dz dk_T^2}(A'_L \to s + A'_L)
=\frac{\alpha'}{2\pi}\frac{z\bar{z}}{2}\left(\frac{m_s^2}{m_{A'}^2}+\frac{2(1-z\bar{z})}{\bar{z}} \right)^2\frac{m_{A'}^2}{\tilde{p}_T^4}.
\end{gather}

\section{Effect of signal distribution on the upper limit} \label{appB}

\hspace{\parindent} In this part, taking $m=0.01$ GeV as an example, we present the detailed process to get the upper limits on the signal yield with and without the dark FSR (denoted as $N'$ and $N_0$, respectively). 

 \begin{figure}[htb]
	\centering
	\includegraphics[width=8.5cm]{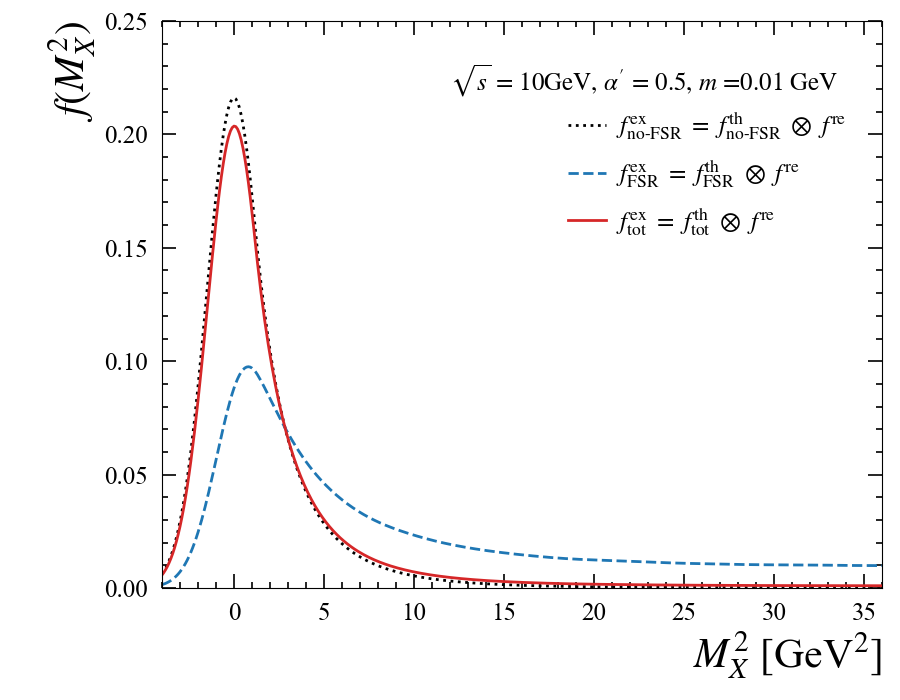}
  \vspace{-.2cm}
  \captionsetup{justification=raggedright, singlelinecheck=false}
	\caption{Probability density functions (PDFs) of $M_X^2$ for signal. The black dotted curve represents the no-FSR scenario, the red solid curve takes the dark FSR into account, and the blue dashed curve corresponds to the component that contributed by the dark FSR.}
\label{fig:sgnpdf}
\end{figure}

Taking the dark FSR into account, the theoretical distribution of $M_X^2$ is composed by $f^{\text{th}}_{\text{no-FSR}}$ and $f^{\text{th}}_{\text{FSR}}$, which correspond to $e^+e^-\to \gamma A'$ and  $e^+e^-\to \gamma A' s$ respectively. 
Considering the detector effect, theoretical distribution should be convolved with detector response function $f^{\text{re}}(M_X^2)$ to obtain the experimentally observed distribution: 
\begin{eqnarray}
f^{\text{ex}}_{\text{no-FSR}}(M_X^2) &=& f^{\text{th}}_{\text{no-FSR}}(M_X^2) \otimes f^{\text{re}}(M_X^2) 
\label{eq:fnofsr}
\end{eqnarray}
\begin{eqnarray}
f^{\text{ex}}_{\text{FSR}}(M_X^2) = f^{\text{th}}_{\text{FSR}}(M_X^2) \otimes f^{\text{re}}(M_X^2).
\label{eq:exFSR}
\end{eqnarray}
\begin{eqnarray}
f^{\text{ex}}_{\text{tot}}(M_X^2) &=& \lambda f^{\text{ex}}_{\text{no-FSR}}(M_X^2)+ (1-\lambda) f^{\text{ex}}_{\text{FSR}}(M_X^2),
\label{eq:fsgn}
\end{eqnarray}
with $\lambda$ the fraction of process $e^+e^-\to \gamma A'$ in the total number of events.

The detector response function $f^{\text{re}}(M_X^2)$ is modeled by a Crystal Ball function with resolution $\sigma(M_X^2) = 1.5~\text{GeV}^2$, and other parameters are determined by fitting the signal peak in Fig.3 of Ref.~\cite{BaBar:2017tiz}.
The $f^{\text{th}}_{\text{FSR}}(M_X^2)$ is obtained via linear interpolation of the histogram from simulated events.
To calculate $f^{\text{ex}}_{\text{FSR}}(M_X^2)$ as Eq.~\ref{eq:exFSR},  we perform a numerical convolution with Fourier transforms in the interval $[-10, 50]~\text{GeV}^2$ using \textsc{RooFit} \cite{Verkerke:2003ir, FFTW05}, with a sampling frequency of 10,000, and the result is shown as the blue dashed line in Fig.~\ref{fig:sgnpdf}. According to Eq.~\ref{eq:fnofsr} and Eq.~\ref{eq:fsgn}, the distributions $f^{\text{ex}}_{\text{no-FSR}}(M_X^2)$ and $f^{\text{ex}}_{\text{tot}}(M_X^2)$ are gotten and indicated by the black dotted line and red line respectively in Fig.~\ref{fig:sgnpdf}.

Notably, as the blue line shows, the radiation of dark Higgs bosons shifts the missing mass squared $M_X^2$ to higher values, flattening the peak, shifting it rightward, and generating a pronounced tail on the high-$M_X^2$ side. However, since only a small fraction of events radiate (no more than $10\%$ for $\alpha'=0.5$), the $f^{\text{ex}}_{\text{tot}}(M_X^2)$ closely resembles the no-FSR distribution,  differing only by a slight reduction in peak height and  a moderate broadening of the tail on the right side.  

\begin{figure}[htb]
	\centering
	\includegraphics[width=8.5cm]{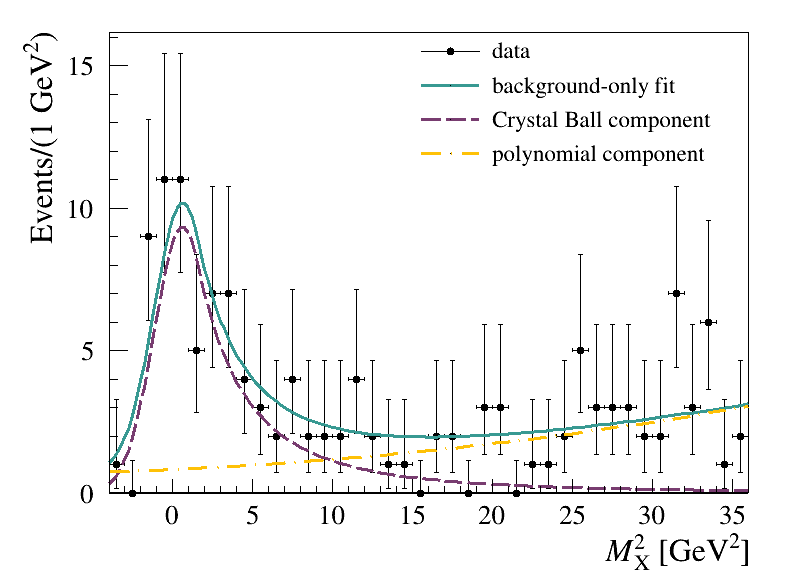}
  \vspace{-.2cm}
 \captionsetup{justification=raggedright, singlelinecheck=false}
	\caption{The distribution of $M_X^2$ in the low-$M_X^2$ region with data samples collected near $\Upsilon(3S)$ resonance and selected with $\mathcal{R}'_L$. Both the histogram and the curve of the background-only fit are taken from the supplementary material of Ref.~\cite{BaBar:2017tiz}.}
\label{fig:babar}
\end{figure} 

To study the impact of changes in signal distribution on experimental results, we perform a simplified recast of the BaBar analysis. There are 6 signal regions for the low-mass situation, and for simplicity, we only focus on the region near the $\Upsilon(3S)$ resonance satisfying the $\mathcal{R}'_L$ selection criteria, which contains 129 events, more than the sum of the other five regions. 
The binned data and the background-only fit are provided in the supplementary material of Ref.~\cite{BaBar:2017tiz}(see Fig.~\ref{fig:babar}). 
We decompose the background-only fit into a Crystal Ball function describing the peaking background from $e^+e^- \to \gamma\gamma$ and a second-order polynomial which primarily comes from Bhabha scattering $e^+e^-\to\gamma e^+e^-$.

With these probability density functions of $M_X^2$, we can construct the profile-likelihood ratio test statistic. Background event counts (total and peaking) are treated as nuisance parameters, and the integral values of the corresponding curves in Fig.~\ref{fig:babar} are set as the global observables. 
Performing hypothesis test inversion using \textsc{RooStats} \cite{Moneta:2010pm, Cowan:2010js} with the binned data, we obtain the $90\%$ confidence level (C.L.) upper limits (based on the CLs method) on the signal yield:  $N_0 = 18.13$ (without dark FSR) and $N' = 18.76$ (with dark FSR). The slight increase in the upper limit arises because incorporating dark FSR leads to a broader signal shape, marginally degrading the experimental sensitivity. 

\addcontentsline{toc}{section}{References}
\bibliographystyle{JHEP}
\bibliography{ref}

\providecommand{\href}[2]{#2}\begingroup\begin{thebibliography}{10}

\bibitem{Abercrombie:2015wmb}
D.~Abercrombie et~al., \textit{{Dark Matter benchmark models for early LHC Run-2 Searches: Report of the ATLAS/CMS Dark Matter Forum}}, \href{https://doi.org/10.1016/j.dark.2019.100371}{\textit{Phys. Dark Univ.} {\bfseries 27} (2020) 100371}, [\href{https://arxiv.org/abs/1507.00966}{{\ttfamily 1507.00966}}].

\bibitem{PandaX-II:2016vec}
{\scshape PandaX-II} collaboration, A.~Tan et~al., \textit{{Dark Matter Results from First 98.7 Days of Data from the PandaX-II Experiment}}, \href{https://doi.org/10.1103/PhysRevLett.117.121303}{\textit{Phys. Rev. Lett.} {\bfseries 117} (2016) 121303}, [\href{https://arxiv.org/abs/1607.07400}{{\ttfamily 1607.07400}}].

\bibitem{LUX:2016ggv}
{\scshape LUX} collaboration, D.~S. Akerib et~al., \textit{{Results from a search for dark matter in the complete LUX exposure}}, \href{https://doi.org/10.1103/PhysRevLett.118.021303}{\textit{Phys. Rev. Lett.} {\bfseries 118} (2017) 021303}, [\href{https://arxiv.org/abs/1608.07648}{{\ttfamily 1608.07648}}].

\bibitem{ATLAS:2017nga}
{\scshape ATLAS} collaboration, M.~Aaboud et~al., \textit{{Search for dark matter at $\sqrt{s}=13$ TeV in final states containing an energetic photon and large missing transverse momentum with the ATLAS detector}}, \href{https://doi.org/10.1140/epjc/s10052-017-4965-8}{\textit{Eur. Phys. J. C} {\bfseries 77} (2017) 393}, [\href{https://arxiv.org/abs/1704.03848}{{\ttfamily 1704.03848}}].

\bibitem{ATLAS:2013ndy}
{\scshape ATLAS} collaboration, G.~Aad et~al., \textit{{Search for dark matter in events with a hadronically decaying W or Z boson and missing transverse momentum in $pp$ collisions at $\sqrt{s} =$ 8 TeV with the ATLAS detector}}, \href{https://doi.org/10.1103/PhysRevLett.112.041802}{\textit{Phys. Rev. Lett.} {\bfseries 112} (2014) 041802}, [\href{https://arxiv.org/abs/1309.4017}{{\ttfamily 1309.4017}}].

\bibitem{CMS:2016gox}
{\scshape CMS} collaboration, V.~Khachatryan et~al., \textit{{Search for dark matter particles in proton-proton collisions at $ \sqrt{s}=8 $ TeV using the razor variables}}, \href{https://doi.org/10.1007/JHEP12(2016)088}{\textit{JHEP} {\bfseries 12} (2016) 088}, [\href{https://arxiv.org/abs/1603.08914}{{\ttfamily 1603.08914}}].

\bibitem{ATLAS:2017uis}
{\scshape ATLAS} collaboration, M.~Aaboud et~al., \textit{{Search for Dark Matter Produced in Association with a Higgs Boson Decaying to $b\bar b$ using 36 fb$^{-1}$ of $pp$ collisions at $\sqrt s=13$ TeV with the ATLAS Detector}}, \href{https://doi.org/10.1103/PhysRevLett.119.181804}{\textit{Phys. Rev. Lett.} {\bfseries 119} (2017) 181804}, [\href{https://arxiv.org/abs/1707.01302}{{\ttfamily 1707.01302}}].

\bibitem{Arkani-Hamed:2008hhe}
N.~Arkani-Hamed, D.~P. Finkbeiner, T.~R. Slatyer and N.~Weiner, \textit{{A Theory of Dark Matter}}, \href{https://doi.org/10.1103/PhysRevD.79.015014}{\textit{Phys. Rev. D} {\bfseries 79} (2009) 015014}, [\href{https://arxiv.org/abs/0810.0713}{{\ttfamily 0810.0713}}].

\bibitem{Pospelov:2008jd}
M.~Pospelov and A.~Ritz, \textit{{Astrophysical Signatures of Secluded Dark Matter}}, \href{https://doi.org/10.1016/j.physletb.2008.12.012}{\textit{Phys. Lett. B} {\bfseries 671} (2009) 391--397}, [\href{https://arxiv.org/abs/0810.1502}{{\ttfamily 0810.1502}}].

\bibitem{Alves:2009nf}
D.~S.~M. Alves, S.~R. Behbahani, P.~Schuster and J.~G. Wacker, \textit{{Composite Inelastic Dark Matter}}, \href{https://doi.org/10.1016/j.physletb.2010.08.006}{\textit{Phys. Lett. B} {\bfseries 692} (2010) 323--326}, [\href{https://arxiv.org/abs/0903.3945}{{\ttfamily 0903.3945}}].

\bibitem{Hisano:2003ec}
J.~Hisano, S.~Matsumoto and M.~M. Nojiri, \textit{{Explosive dark matter annihilation}}, \href{https://doi.org/10.1103/PhysRevLett.92.031303}{\textit{Phys. Rev. Lett.} {\bfseries 92} (2004) 031303}, [\href{https://arxiv.org/abs/hep-ph/0307216}{{\ttfamily hep-ph/0307216}}].

\bibitem{March-Russell:2008lng}
J.~March-Russell, S.~M. West, D.~Cumberbatch and D.~Hooper, \textit{{Heavy Dark Matter Through the Higgs Portal}}, \href{https://doi.org/10.1088/1126-6708/2008/07/058}{\textit{JHEP} {\bfseries 07} (2008) 058}, [\href{https://arxiv.org/abs/0801.3440}{{\ttfamily 0801.3440}}].

\bibitem{Arkani-Hamed:2008kxc}
N.~Arkani-Hamed and N.~Weiner, \textit{{LHC Signals for a SuperUnified Theory of Dark Matter}}, \href{https://doi.org/10.1088/1126-6708/2008/12/104}{\textit{JHEP} {\bfseries 12} (2008) 104}, [\href{https://arxiv.org/abs/0810.0714}{{\ttfamily 0810.0714}}].

\bibitem{Cirelli:2008pk}
M.~Cirelli, M.~Kadastik, M.~Raidal and A.~Strumia, \textit{{Model-independent implications of the e+-, anti-proton cosmic ray spectra on properties of Dark Matter}}, \href{https://doi.org/10.1016/j.nuclphysb.2008.11.031}{\textit{Nucl. Phys. B} {\bfseries 813} (2009) 1--21}, [\href{https://arxiv.org/abs/0809.2409}{{\ttfamily 0809.2409}}]. [Addendum: Nucl.Phys.B 873, 530--533 (2013)].

\bibitem{Pospelov:2007mp}
M.~Pospelov, A.~Ritz and M.~B. Voloshin, \textit{{Secluded WIMP Dark Matter}}, \href{https://doi.org/10.1016/j.physletb.2008.02.052}{\textit{Phys. Lett. B} {\bfseries 662} (2008) 53--61}, [\href{https://arxiv.org/abs/0711.4866}{{\ttfamily 0711.4866}}].

\bibitem{Cholis:2008wq}
I.~Cholis, G.~Dobler, D.~P. Finkbeiner, L.~Goodenough and N.~Weiner, \textit{{The Case for a 700+ GeV WIMP: Cosmic Ray Spectra from ATIC and PAMELA}}, \href{https://doi.org/10.1103/PhysRevD.80.123518}{\textit{Phys. Rev. D} {\bfseries 80} (2009) 123518}, [\href{https://arxiv.org/abs/0811.3641}{{\ttfamily 0811.3641}}].

\bibitem{Cholis:2008qq}
I.~Cholis, D.~P. Finkbeiner, L.~Goodenough and N.~Weiner, \textit{{The PAMELA Positron Excess from Annihilations into a Light Boson}}, \href{https://doi.org/10.1088/1475-7516/2009/12/007}{\textit{JCAP} {\bfseries 12} (2009) 007}, [\href{https://arxiv.org/abs/0810.5344}{{\ttfamily 0810.5344}}].

\bibitem{Rizzo:2024bhn}
T.~G. Rizzo, \textit{{Toward UV models of kinetic mixing and portal matter. VI. A more complex dark matter sector?}}, \href{https://doi.org/10.1103/PhysRevD.110.075037}{\textit{Phys. Rev. D} {\bfseries 110} (2024) 075037}, [\href{https://arxiv.org/abs/2408.01296}{{\ttfamily 2408.01296}}].

\bibitem{SHiP:2021nfo}
{\scshape SHiP} collaboration, C.~Ahdida et~al., \textit{{The SHiP experiment at the proposed CERN SPS Beam Dump Facility}}, \href{https://doi.org/10.1140/epjc/s10052-022-10346-5}{\textit{Eur. Phys. J. C} {\bfseries 82} (2022) 486}, [\href{https://arxiv.org/abs/2112.01487}{{\ttfamily 2112.01487}}].

\bibitem{BaBar:2001yhh}
{\scshape BaBar} collaboration, B.~Aubert et~al., \textit{{The BaBar detector}}, \href{https://doi.org/10.1016/S0168-9002(01)02012-5}{\textit{Nucl. Instrum. Meth. A} {\bfseries 479} (2002) 1--116}, [\href{https://arxiv.org/abs/hep-ex/0105044}{{\ttfamily hep-ex/0105044}}].

\bibitem{Belle:2000cnh}
{\scshape Belle} collaboration, A.~Abashian et~al., \textit{{The Belle Detector}}, \href{https://doi.org/10.1016/S0168-9002(01)02013-7}{\textit{Nucl. Instrum. Meth. A} {\bfseries 479} (2002) 117--232}.

\bibitem{BESIII:2009fln}
{\scshape BESIII} collaboration, M.~Ablikim et~al., \textit{{Design and Construction of the BESIII Detector}}, \href{https://doi.org/10.1016/j.nima.2009.12.050}{\textit{Nucl. Instrum. Meth. A} {\bfseries 614} (2010) 345--399}, [\href{https://arxiv.org/abs/0911.4960}{{\ttfamily 0911.4960}}].

\bibitem{Adinolfi:2002uk}
M.~Adinolfi et~al., \textit{{The tracking detector of the KLOE experiment}}, \href{https://doi.org/10.1016/S0168-9002(02)00514-4}{\textit{Nucl. Instrum. Meth. A} {\bfseries 488} (2002) 51--73}.

\bibitem{Andreas:2013lya}
S.~Andreas et~al., \textit{{Proposal for an Experiment to Search for Light Dark Matter at the SPS}},  \href{https://arxiv.org/abs/1312.3309}{{\ttfamily 1312.3309}}.

\bibitem{NA64:2016oww}
{\scshape NA64} collaboration, D.~Banerjee et~al., \textit{{Search for invisible decays of sub-GeV dark photons in missing-energy events at the CERN SPS}}, \href{https://doi.org/10.1103/PhysRevLett.118.011802}{\textit{Phys. Rev. Lett.} {\bfseries 118} (2017) 011802}, [\href{https://arxiv.org/abs/1610.02988}{{\ttfamily 1610.02988}}].

\bibitem{Gninenko:2013rka}
S.~N. Gninenko, \textit{{Search for MeV dark photons in a light-shining-through-walls experiment at CERN}}, \href{https://doi.org/10.1103/PhysRevD.89.075008}{\textit{Phys. Rev. D} {\bfseries 89} (2014) 075008}, [\href{https://arxiv.org/abs/1308.6521}{{\ttfamily 1308.6521}}].

\bibitem{NA64:2017vtt}
{\scshape NA64} collaboration, D.~Banerjee et~al., \textit{{Search for vector mediator of Dark Matter production in invisible decay mode}}, \href{https://doi.org/10.1103/PhysRevD.97.072002}{\textit{Phys. Rev. D} {\bfseries 97} (2018) 072002}, [\href{https://arxiv.org/abs/1710.00971}{{\ttfamily 1710.00971}}].

\bibitem{DarkSHINE:2022mak}
{\scshape DarkSHINE} collaboration, J.~Chen et~al., \textit{{Prospective study of light dark matter search with a newly proposed DarkSHINE experiment}}, \href{https://doi.org/10.1007/s11433-022-1983-8}{\textit{Sci. China Phys. Mech. Astron.} {\bfseries 66} (2023) 211062}.

\bibitem{Yan:2022npz}
B.~Yan, \textit{{Probing the dark photon via polarized DIS scattering at the HERA and EIC}}, \href{https://doi.org/10.1016/j.physletb.2022.137384}{\textit{Phys. Lett. B} {\bfseries 833} (2022) 137384}, [\href{https://arxiv.org/abs/2203.01510}{{\ttfamily 2203.01510}}].

\bibitem{Feng:2017uoz}
J.~L. Feng, I.~Galon, F.~Kling and S.~Trojanowski, \textit{{ForwArd Search ExpeRiment at the LHC}}, \href{https://doi.org/10.1103/PhysRevD.97.035001}{\textit{Phys. Rev. D} {\bfseries 97} (2018) 035001}, [\href{https://arxiv.org/abs/1708.09389}{{\ttfamily 1708.09389}}].

\bibitem{FASER:2022hcn}
{\scshape FASER} collaboration, H.~Abreu et~al., \textit{{The FASER detector}}, \href{https://doi.org/10.1088/1748-0221/19/05/P05066}{\textit{JINST} {\bfseries 19} (2024) P05066}, [\href{https://arxiv.org/abs/2207.11427}{{\ttfamily 2207.11427}}].

\bibitem{LHCb:2008vvz}
{\scshape LHCb} collaboration, A.~A. Alves, Jr. et~al., \textit{{The LHCb Detector at the LHC}}, \href{https://doi.org/10.1088/1748-0221/3/08/S08005}{\textit{JINST} {\bfseries 3} (2008) S08005}.

\bibitem{Banerjee:2019pds}
D.~Banerjee et~al., \textit{{Dark matter search in missing energy events with NA64}}, \href{https://doi.org/10.1103/PhysRevLett.123.121801}{\textit{Phys. Rev. Lett.} {\bfseries 123} (2019) 121801}, [\href{https://arxiv.org/abs/1906.00176}{{\ttfamily 1906.00176}}].

\bibitem{Zhang:2019wnz}
Y.~Zhang, W.-T. Zhang, M.~Song, X.-A. Pan, Z.-M. Niu and G.~Li, \textit{{Probing invisible decay of dark photon at BESIII and future STCF via monophoton searches}}, \href{https://doi.org/10.1103/PhysRevD.100.115016}{\textit{Phys. Rev. D} {\bfseries 100} (2019) 115016}, [\href{https://arxiv.org/abs/1907.07046}{{\ttfamily 1907.07046}}].

\bibitem{BaBar:2017tiz}
{\scshape BaBar} collaboration, J.~P. Lees et~al., \textit{{Search for Invisible Decays of a Dark Photon Produced in ${e}^{+}{e}^{-}$ Collisions at BaBar}}, \href{https://doi.org/10.1103/PhysRevLett.119.131804}{\textit{Phys. Rev. Lett.} {\bfseries 119} (2017) 131804}, [\href{https://arxiv.org/abs/1702.03327}{{\ttfamily 1702.03327}}].

\bibitem{BESIII:2017fwv}
{\scshape BESIII} collaboration, M.~Ablikim et~al., \textit{{Dark Photon Search in the Mass Range Between 1.5 and 3.4 GeV/$c^2$}}, \href{https://doi.org/10.1016/j.physletb.2017.09.067}{\textit{Phys. Lett. B} {\bfseries 774} (2017) 252--257}, [\href{https://arxiv.org/abs/1705.04265}{{\ttfamily 1705.04265}}].

\bibitem{Ilten:2015hya}
P.~Ilten, J.~Thaler, M.~Williams and W.~Xue, \textit{{Dark photons from charm mesons at LHCb}}, \href{https://doi.org/10.1103/PhysRevD.92.115017}{\textit{Phys. Rev. D} {\bfseries 92} (2015) 115017}, [\href{https://arxiv.org/abs/1509.06765}{{\ttfamily 1509.06765}}].

\bibitem{KLOE-2:2012lii}
{\scshape KLOE-2} collaboration, D.~Babusci et~al., \textit{{Limit on the production of a light vector gauge boson in phi meson decays with the KLOE detector}}, \href{https://doi.org/10.1016/j.physletb.2013.01.067}{\textit{Phys. Lett. B} {\bfseries 720} (2013) 111--115}, [\href{https://arxiv.org/abs/1210.3927}{{\ttfamily 1210.3927}}].

\bibitem{KLOE-2:2016ydq}
{\scshape KLOE-2} collaboration, A.~Anastasi et~al., \textit{{Limit on the production of a new vector boson in $\mathrm{e^+ e^-}\rightarrow {\rm U}\gamma$, U$\rightarrow \pi^+\pi^-$ with the KLOE experiment}}, \href{https://doi.org/10.1016/j.physletb.2016.04.019}{\textit{Phys. Lett. B} {\bfseries 757} (2016) 356--361}, [\href{https://arxiv.org/abs/1603.06086}{{\ttfamily 1603.06086}}].

\bibitem{Ilten:2016tkc}
P.~Ilten, Y.~Soreq, J.~Thaler, M.~Williams and W.~Xue, \textit{{Proposed Inclusive Dark Photon Search at LHCb}}, \href{https://doi.org/10.1103/PhysRevLett.116.251803}{\textit{Phys. Rev. Lett.} {\bfseries 116} (2016) 251803}, [\href{https://arxiv.org/abs/1603.08926}{{\ttfamily 1603.08926}}].

\bibitem{Merkel:2014avp}
H.~Merkel et~al., \textit{{Search at the Mainz Microtron for Light Massive Gauge Bosons Relevant for the Muon g-2 Anomaly}}, \href{https://doi.org/10.1103/PhysRevLett.112.221802}{\textit{Phys. Rev. Lett.} {\bfseries 112} (2014) 221802}, [\href{https://arxiv.org/abs/1404.5502}{{\ttfamily 1404.5502}}].

\bibitem{BaBar:2014zli}
{\scshape BaBar} collaboration, J.~P. Lees et~al., \textit{{Search for a Dark Photon in $e^+e^-$ Collisions at BaBar}}, \href{https://doi.org/10.1103/PhysRevLett.113.201801}{\textit{Phys. Rev. Lett.} {\bfseries 113} (2014) 201801}, [\href{https://arxiv.org/abs/1406.2980}{{\ttfamily 1406.2980}}].

\bibitem{KLOE-2:2011hhj}
{\scshape KLOE-2} collaboration, F.~Archilli et~al., \textit{{Search for a vector gauge boson in $\phi$ meson decays with the KLOE detector}}, \href{https://doi.org/10.1016/j.physletb.2011.11.033}{\textit{Phys. Lett. B} {\bfseries 706} (2012) 251--255}, [\href{https://arxiv.org/abs/1110.0411}{{\ttfamily 1110.0411}}].

\bibitem{LHCb:2019vmc}
{\scshape LHCb} collaboration, R.~Aaij et~al., \textit{{Search for $A'\to\mu^+\mu^-$ Decays}}, \href{https://doi.org/10.1103/PhysRevLett.124.041801}{\textit{Phys. Rev. Lett.} {\bfseries 124} (2020) 041801}, [\href{https://arxiv.org/abs/1910.06926}{{\ttfamily 1910.06926}}].

\bibitem{KLOE-2:2014qxg}
{\scshape KLOE-2} collaboration, D.~Babusci et~al., \textit{{Search for light vector boson production in $e^+e^- \rightarrow \mu^+ \mu^- \gamma$ interactions with the KLOE experiment}}, \href{https://doi.org/10.1016/j.physletb.2014.08.005}{\textit{Phys. Lett. B} {\bfseries 736} (2014) 459--464}, [\href{https://arxiv.org/abs/1404.7772}{{\ttfamily 1404.7772}}].

\bibitem{NA482:2015wmo}
{\scshape NA48/2} collaboration, J.~R. Batley et~al., \textit{{Search for the dark photon in $\pi^0$ decays}}, \href{https://doi.org/10.1016/j.physletb.2015.04.068}{\textit{Phys. Lett. B} {\bfseries 746} (2015) 178--185}, [\href{https://arxiv.org/abs/1504.00607}{{\ttfamily 1504.00607}}].

\bibitem{Belle-II:2018jsg}
{\scshape Belle-II} collaboration, W.~Altmannshofer et~al., \textit{{The Belle II Physics Book}}, \href{https://doi.org/10.1093/ptep/ptz106}{\textit{PTEP} {\bfseries 2019} (2019) 123C01}, [\href{https://arxiv.org/abs/1808.10567}{{\ttfamily 1808.10567}}]. [Erratum: PTEP 2020, 029201 (2020)].

\bibitem{Lee:1977yc}
B.~W. Lee, C.~Quigg and H.~B. Thacker, \textit{{The Strength of Weak Interactions at Very High-Energies and the Higgs Boson Mass}}, \href{https://doi.org/10.1103/PhysRevLett.38.883}{\textit{Phys. Rev. Lett.} {\bfseries 38} (1977) 883--885}.

\bibitem{Lee:1977eg}
B.~W. Lee, C.~Quigg and H.~B. Thacker, \textit{{Weak Interactions at Very High-Energies: The Role of the Higgs Boson Mass}}, \href{https://doi.org/10.1103/PhysRevD.16.1519}{\textit{Phys. Rev. D} {\bfseries 16} (1977) 1519}.

\bibitem{Li:2024wqj}
S.~Li, J.~M. Yang, M.~Zhang and R.~Zhu, \textit{{Theoretical bounds on dark Higgs mass in a self-interacting dark matter model with U(1)'}}, \href{https://doi.org/10.1103/PhysRevD.111.035005}{\textit{Phys. Rev. D} {\bfseries 111} (2025) 035005}, [\href{https://arxiv.org/abs/2405.18226}{{\ttfamily 2405.18226}}].

\bibitem{Chen:2023rrl}
Z.~Chen, K.~Ye and M.~Zhang, \textit{{Asymmetric dark matter with a spontaneously broken U(1)': Self-interaction and gravitational waves}}, \href{https://doi.org/10.1103/PhysRevD.107.095027}{\textit{Phys. Rev. D} {\bfseries 107} (2023) 095027}, [\href{https://arxiv.org/abs/2303.11820}{{\ttfamily 2303.11820}}].

\bibitem{Baldes:2017gzw}
I.~Baldes and K.~Petraki, \textit{{Asymmetric thermal-relic dark matter: Sommerfeld-enhanced freeze-out, annihilation signals and unitarity bounds}}, \href{https://doi.org/10.1088/1475-7516/2017/09/028}{\textit{JCAP} {\bfseries 09} (2017) 028}, [\href{https://arxiv.org/abs/1703.00478}{{\ttfamily 1703.00478}}].

\bibitem{Graesser:2011wi}
M.~L. Graesser, I.~M. Shoemaker and L.~Vecchi, \textit{{Asymmetric WIMP dark matter}}, \href{https://doi.org/10.1007/JHEP10(2011)110}{\textit{JHEP} {\bfseries 10} (2011) 110}, [\href{https://arxiv.org/abs/1103.2771}{{\ttfamily 1103.2771}}].

\bibitem{Ibe:2019ena}
M.~Ibe, A.~Kamada, S.~Kobayashi, T.~Kuwahara and W.~Nakano, \textit{{Baryon-Dark Matter Coincidence in Mirrored Unification}}, \href{https://doi.org/10.1103/PhysRevD.100.075022}{\textit{Phys. Rev. D} {\bfseries 100} (2019) 075022}, [\href{https://arxiv.org/abs/1907.03404}{{\ttfamily 1907.03404}}].

\bibitem{Batell:2009yf}
B.~Batell, M.~Pospelov and A.~Ritz, \textit{{Probing a Secluded U(1) at B-factories}}, \href{https://doi.org/10.1103/PhysRevD.79.115008}{\textit{Phys. Rev. D} {\bfseries 79} (2009) 115008}, [\href{https://arxiv.org/abs/0903.0363}{{\ttfamily 0903.0363}}].

\bibitem{Cheung:2024oxh}
K.~Cheung, Y.~Kim, Y.~Kwon, C.~J. Ouseph, A.~Soffer and Z.~S. Wang, \textit{{Probing dark photons from a light scalar at Belle II}}, \href{https://doi.org/10.1007/JHEP05(2024)094}{\textit{JHEP} {\bfseries 05} (2024) 094}, [\href{https://arxiv.org/abs/2401.03168}{{\ttfamily 2401.03168}}].

\bibitem{Belle-II:2022jyy}
{\scshape Belle-II} collaboration, F.~Abudin\'en et~al., \textit{{Search for a Dark Photon and an Invisible Dark Higgs Boson in \ensuremath{\mu}+\ensuremath{\mu}- and Missing Energy Final States with the Belle II Experiment}}, \href{https://doi.org/10.1103/PhysRevLett.130.071804}{\textit{Phys. Rev. Lett.} {\bfseries 130} (2023) 071804}, [\href{https://arxiv.org/abs/2207.00509}{{\ttfamily 2207.00509}}].

\bibitem{KLOE-2:2015nli}
{\scshape KLOE-2} collaboration, A.~Anastasi et~al., \textit{{Search for dark Higgsstrahlung in $e^{+}e^{-} \to \mu^{+}\mu^-$ and missing energy events with the KLOE experiment}}, \href{https://doi.org/10.1016/j.physletb.2015.06.015}{\textit{Phys. Lett. B} {\bfseries 747} (2015) 365--372}, [\href{https://arxiv.org/abs/1501.06795}{{\ttfamily 1501.06795}}].

\bibitem{BaBar:2012bkw}
{\scshape BaBar} collaboration, J.~P. Lees et~al., \textit{{Search for Low-Mass Dark-Sector Higgs Bosons}}, \href{https://doi.org/10.1103/PhysRevLett.108.211801}{\textit{Phys. Rev. Lett.} {\bfseries 108} (2012) 211801}, [\href{https://arxiv.org/abs/1202.1313}{{\ttfamily 1202.1313}}].

\bibitem{Jaegle:2015fme}
{\scshape Belle} collaboration, I.~Jaegle, \textit{{Search for the dark photon and the dark Higgs boson at Belle}}, \href{https://doi.org/10.1103/PhysRevLett.114.211801}{\textit{Phys. Rev. Lett.} {\bfseries 114} (2015) 211801}, [\href{https://arxiv.org/abs/1502.00084}{{\ttfamily 1502.00084}}].

\bibitem{Buschmann:2015awa}
M.~Buschmann, J.~Kopp, J.~Liu and P.~A.~N. Machado, \textit{{Lepton Jets from Radiating Dark Matter}}, \href{https://doi.org/10.1007/JHEP07(2015)045}{\textit{JHEP} {\bfseries 07} (2015) 045}, [\href{https://arxiv.org/abs/1505.07459}{{\ttfamily 1505.07459}}].

\bibitem{Kim:2016fdv}
M.~Kim, H.-S. Lee, M.~Park and M.~Zhang, \textit{{Examining the origin of dark matter mass at colliders}}, \href{https://doi.org/10.1103/PhysRevD.98.055027}{\textit{Phys. Rev. D} {\bfseries 98} (2018) 055027}, [\href{https://arxiv.org/abs/1612.02850}{{\ttfamily 1612.02850}}].

\bibitem{Chen:2018uii}
J.~Chen, P.~Ko, H.-N. Li, J.~Li and H.~Yokoya, \textit{{Light dark matter showering under broken dark $U(1)$ \textemdash{} revisited}}, \href{https://doi.org/10.1007/JHEP01(2019)141}{\textit{JHEP} {\bfseries 01} (2019) 141}, [\href{https://arxiv.org/abs/1807.00530}{{\ttfamily 1807.00530}}].

\bibitem{Chigusa:2022act}
S.~Chigusa and M.~Yamazaki, \textit{{Quantum simulations of dark sector showers}}, \href{https://doi.org/10.1016/j.physletb.2022.137466}{\textit{Phys. Lett. B} {\bfseries 834} (2022) 137466}, [\href{https://arxiv.org/abs/2204.12500}{{\ttfamily 2204.12500}}].

\bibitem{Bengtsson:1986et}
M.~Bengtsson and T.~Sjostrand, \textit{{A Comparative Study of Coherent and Noncoherent Parton Shower Evolution}}, \href{https://doi.org/10.1016/0550-3213(87)90407-X}{\textit{Nucl. Phys. B} {\bfseries 289} (1987) 810--846}.

\bibitem{Chen:2016wkt}
J.~Chen, T.~Han and B.~Tweedie, \textit{{Electroweak Splitting Functions and High Energy Showering}}, \href{https://doi.org/10.1007/JHEP11(2017)093}{\textit{JHEP} {\bfseries 11} (2017) 093}, [\href{https://arxiv.org/abs/1611.00788}{{\ttfamily 1611.00788}}].

\bibitem{Collins:1984kg}
J.~C. Collins, D.~E. Soper and G.~F. Sterman, \textit{{Transverse Momentum Distribution in Drell-Yan Pair and W and Z Boson Production}}, \href{https://doi.org/10.1016/0550-3213(85)90479-1}{\textit{Nucl. Phys. B} {\bfseries 250} (1985) 199--224}.

\bibitem{Collins:1989gx}
J.~C. Collins, D.~E. Soper and G.~F. Sterman, \textit{{Factorization of Hard Processes in QCD}}, \href{https://doi.org/10.1142/9789814503266_0001}{\textit{Adv. Ser. Direct. High Energy Phys.} {\bfseries 5} (1989) 1--91}, [\href{https://arxiv.org/abs/hep-ph/0409313}{{\ttfamily hep-ph/0409313}}].

\bibitem{Sudakov:1954sw}
V.~V. Sudakov, \textit{{Vertex parts at very high-energies in quantum electrodynamics}}, {\textit{Sov. Phys. JETP} {\bfseries 3} (1956) 65--71}.

\bibitem{Hoeche:2009xc}
S.~Hoeche, S.~Schumann and F.~Siegert, \textit{{Hard photon production and matrix-element parton-shower merging}}, \href{https://doi.org/10.1103/PhysRevD.81.034026}{\textit{Phys. Rev. D} {\bfseries 81} (2010) 034026}, [\href{https://arxiv.org/abs/0912.3501}{{\ttfamily 0912.3501}}].

\bibitem{Lonnblad:2012hz}
L.~L\"onnblad, \textit{{Fooling Around with the Sudakov Veto Algorithm}}, \href{https://doi.org/10.1140/epjc/s10052-013-2350-9}{\textit{Eur. Phys. J. C} {\bfseries 73} (2013) 2350}, [\href{https://arxiv.org/abs/1211.7204}{{\ttfamily 1211.7204}}].

\bibitem{Mrenna:2016sih}
S.~Mrenna and P.~Skands, \textit{{Automated Parton-Shower Variations in Pythia 8}}, \href{https://doi.org/10.1103/PhysRevD.94.074005}{\textit{Phys. Rev. D} {\bfseries 94} (2016) 074005}, [\href{https://arxiv.org/abs/1605.08352}{{\ttfamily 1605.08352}}].

\bibitem{Bierlich:2022pfr}
C.~Bierlich et~al., \textit{{A comprehensive guide to the physics and usage of PYTHIA 8.3}},  \href{https://arxiv.org/abs/2203.11601}{{\ttfamily 2203.11601}}.

\bibitem{Carloni:2010tw}
L.~Carloni and T.~Sjostrand, \textit{{Visible Effects of Invisible Hidden Valley Radiation}}, \href{https://doi.org/10.1007/JHEP09(2010)105}{\textit{JHEP} {\bfseries 09} (2010) 105}, [\href{https://arxiv.org/abs/1006.2911}{{\ttfamily 1006.2911}}].

\bibitem{Carloni:2011kk}
L.~Carloni, J.~Rathsman and T.~Sjostrand, \textit{{Discerning Secluded Sector gauge structures}}, \href{https://doi.org/10.1007/JHEP04(2011)091}{\textit{JHEP} {\bfseries 04} (2011) 091}, [\href{https://arxiv.org/abs/1102.3795}{{\ttfamily 1102.3795}}].

\bibitem{Alloul:2013bka}
A.~Alloul, N.~D. Christensen, C.~Degrande, C.~Duhr and B.~Fuks, \textit{{FeynRules 2.0 - A complete toolbox for tree-level phenomenology}}, \href{https://doi.org/10.1016/j.cpc.2014.04.012}{\textit{Comput. Phys. Commun.} {\bfseries 185} (2014) 2250--2300}, [\href{https://arxiv.org/abs/1310.1921}{{\ttfamily 1310.1921}}].

\bibitem{Alwall:2014hca}
J.~Alwall, R.~Frederix, S.~Frixione, V.~Hirschi, F.~Maltoni, O.~Mattelaer et~al., \textit{{The automated computation of tree-level and next-to-leading order differential cross sections, and their matching to parton shower simulations}}, \href{https://doi.org/10.1007/JHEP07(2014)079}{\textit{JHEP} {\bfseries 07} (2014) 079}, [\href{https://arxiv.org/abs/1405.0301}{{\ttfamily 1405.0301}}].

\bibitem{Frederix:2018nkq}
R.~Frederix, S.~Frixione, V.~Hirschi, D.~Pagani, H.~S. Shao and M.~Zaro, \textit{{The automation of next-to-leading order electroweak calculations}}, \href{https://doi.org/10.1007/JHEP11(2021)085}{\textit{JHEP} {\bfseries 07} (2018) 185}, [\href{https://arxiv.org/abs/1804.10017}{{\ttfamily 1804.10017}}]. [Erratum: JHEP 11, 085 (2021)].

\bibitem{Dev:2024twk}
P.~S.~B. Dev, D.~Kim, D.~Sathyan, K.~Sinha and Y.~Zhang, \textit{{New Laboratory Constraints on Neutrinophilic Mediators}},  \href{https://arxiv.org/abs/2407.12738}{{\ttfamily 2407.12738}}.

\bibitem{Kinoshita:1962ur}
T.~Kinoshita, \textit{{Mass singularities of Feynman amplitudes}}, \href{https://doi.org/10.1063/1.1724268}{\textit{J. Math. Phys.} {\bfseries 3} (1962) 650--677}.

\bibitem{Lee:1964is}
T.~D. Lee and M.~Nauenberg, \textit{{Degenerate Systems and Mass Singularities}}, \href{https://doi.org/10.1103/PhysRev.133.B1549}{\textit{Phys. Rev.} {\bfseries 133} (1964) B1549--B1562}.

\bibitem{Moneta:2010pm}
L.~Moneta, K.~Belasco, K.~S. Cranmer, S.~Kreiss, A.~Lazzaro, D.~Piparo et~al., \textit{{The RooStats Project}}, \href{https://doi.org/10.22323/1.093.0057}{\textit{PoS} {\bfseries ACAT2010} (2010) 057}, [\href{https://arxiv.org/abs/1009.1003}{{\ttfamily 1009.1003}}].

\bibitem{Cowan:2010js}
G.~Cowan, K.~Cranmer, E.~Gross and O.~Vitells, \textit{{Asymptotic formulae for likelihood-based tests of new physics}}, \href{https://doi.org/10.1140/epjc/s10052-011-1554-0}{\textit{Eur. Phys. J. C} {\bfseries 71} (2011) 1554}, [\href{https://arxiv.org/abs/1007.1727}{{\ttfamily 1007.1727}}]. [Erratum: Eur.Phys.J.C 73, 2501 (2013)].

\bibitem{Verkerke:2003ir}
W.~Verkerke and D.~P. Kirkby, \textit{{The RooFit toolkit for data modeling}}, {\textit{eConf} {\bfseries C0303241} (2003) MOLT007}, [\href{https://arxiv.org/abs/physics/0306116}{{\ttfamily physics/0306116}}].

\bibitem{FFTW05}
M.~Frigo and S.~G. Johnson, \textit{The design and implementation of {FFTW3}}, {\textit{Proceedings of the IEEE} {\bfseries 93} (2005) 216--231}. Special issue on ``Program Generation, Optimization, and Platform Adaptation''.

\end{thebibliography}\endgroup

\end{document}